\documentclass[aps,prl,reprint,superscriptaddress,amsmath,amssymb,stfloats]{revtex4-1}

\usepackage{textcomp,SIunits}
\usepackage{graphicx}
\usepackage{dcolumn}
\usepackage{bm}
\usepackage{flafter}

\begin{document}

\title{Closed-loop Control of Compensation Point in the K-Rb-$^{21}$Ne Comagnetometer}

\author{Liwei Jiang}
\affiliation{ School of Instrumentation and Optoelectronic Engineering, Beihang University, Beijing 100191, China}
\author{Wei Quan}
\email[]{quanwei@buaa.edu.cn}
\affiliation{ School of Instrumentation and Optoelectronic Engineering, Beihang University, Beijing 100191, China}
\affiliation{Science and Technology on Inertial Laboratory, Beihang University, Beijing 100191, China}
\author{Feng Liu}
\affiliation{ School of Instrumentation and Optoelectronic Engineering, Beihang University, Beijing 100191, China}
\author{Wenfeng Fan}
\affiliation{ School of Instrumentation and Optoelectronic Engineering, Beihang University, Beijing 100191, China}

\author{Li Xing}
\affiliation{ School of Instrumentation and Optoelectronic Engineering, Beihang University, Beijing 100191, China}
\author{Lihong Duan}
\affiliation{ School of Instrumentation and Optoelectronic Engineering, Beihang University, Beijing 100191, China}
\author{Wuming Liu}
\affiliation{ Beijing National Laboratory for Condensed Matter Physics, Institute of Physics, Chinese Academy of Sciences, Beijing, 100190, China}
\affiliation{ School of Physical Sciences, University of Chinese Academy of Sciences, Beijing, 100190, China}

\author{Jiancheng Fang}
\affiliation{ School of Instrumentation and Optoelectronic Engineering, Beihang University, Beijing 100191, China}
\affiliation{Science and Technology on Inertial Laboratory, Beihang University, Beijing 100191, China}

\date{\today}

\begin{abstract}

We investigate the real-time closed-loop control of compensation point in the K-Rb-$^{21}$Ne comagnetometer operated in the spin-exchange relaxation-free regime. By locking the electron resonance, the alkali metal electrons are free from the fluctuations of the longitudinal ambient magnetic field and nuclear magnetization, which could improve the systematic stability, enlarge the linear measuring range, and suppress the cross-talk error of the comagnetometer. This is the first demonstration of closed-loop control of magnetic field in the single nuclear species comagnetometer, which will be of great significance for rotation sensing as gyroscopes and other high precision metrology applications of the comagnetometer.

\end{abstract}

\maketitle

Atomic comagnetometers, which use at least two spin species to measure magnetic fields in the same space and time, have found a wide range of applications, such as tests of CPT and Lorentz invariance \cite{Limit2000,datatableCPT,KRbNe2011,CPT2014}, searches for anomalous spin-dependent forces \cite{spinforce2009,spinForce2013,Science,PRD}, and inertial rotation sensing \cite{kornack2005,donley2010,He3Xe129,Jiang2018,zhangC}. In all of these applications, the long-term stability of the comagnetometer is essential and often limited by noise and systematic effects associated with the external magnetic fields and magnetization due to spin dipolar interactions \cite{Meg,ShengD}. In general, comagnetometers with two or more nuclear species calculate an appropriate combination of nuclear precession frequencies to cancel out magnetic field dependence \cite{comag,kanegsberg2007,2016limit}. Based on dual nuclear isotopes differential technique, a three-axis residual magnetic fields closed-loop control system has been incorporated in NMR gyros to guarantee the long-term stability of magnetic fields \cite{grover1979,Meyer2014,WalkerNMRG}.

 However, it is a challenging and almost unexplored topic to control magnetic field fluctuations in the single nuclear species comagnetometers. The primary difficulty is acquiring magnetic field information as the feedback signal without the exterior sensors. To circumvent spin precession due to magnetic fields as well as their gradients, the spin-exchange relaxation-free (SERF) comagnetometer involving alkali metals and one kind of nuclear species was first introduced in Ref.~\cite{kornack2002}. A bias magnetic field parallel to the pump laser beam, which is referred as compensation point, cancels the fields from electron and nuclear magnetization and operates the atomic spins in a self-compensating regime, where the nuclear magnetization adiabatically follows slow changes in the external magnetic field, decreasing the effect of transverse fields on alkali metal electron spins \cite{kornack2005,Brown,ChenYdynamics}. Despite the sensitivity to magnetic fields has been suppressed in the SERF comagnetometer, the drifts of external magnetic fields still arise a significant influence on the systematic stability \cite{LI2017}. Moreover, the fluctuation of compensation point will also cause a cross-talk error in the dual-axis SERF comagnetometer \cite{jiang2017,ChenSpin}. The applications of the SERF comagnetometer confront considerable obstacles due to the uncontrolled compensation magnetic field in the system.

In this Letter, we demonstrate a real-time closed-loop control method to stabilize the compensation point of K-Rb-$^{21}$Ne comagnetometer. We find that the electron resonance is shifted to high frequency and separated clearly from the nuclear resonance by the large field from electron magnetization in the K-Rb-$^{21}$Ne comagnetometer. The electron resonance frequency and phase scale with the shift of the compensate point, which allows us to achieve the closed-loop control of the compensation point by locking the electron resonance. This method is validated theoretically and experimentally in our dual-axis K-Rb-$^{21}$Ne comagnetometer. With the closed-loop control of the compensation point, the alkali metal electrons are immune to the fluctuations of the longitudinal ambient magnetic field and nuclear magnetization, which could improve the systematic stability, enlarge the linear measuring range, and suppress the cross-talk error of the comagnetometer.

\begin{figure}[t]
\begin{center}
\includegraphics[width=8cm]{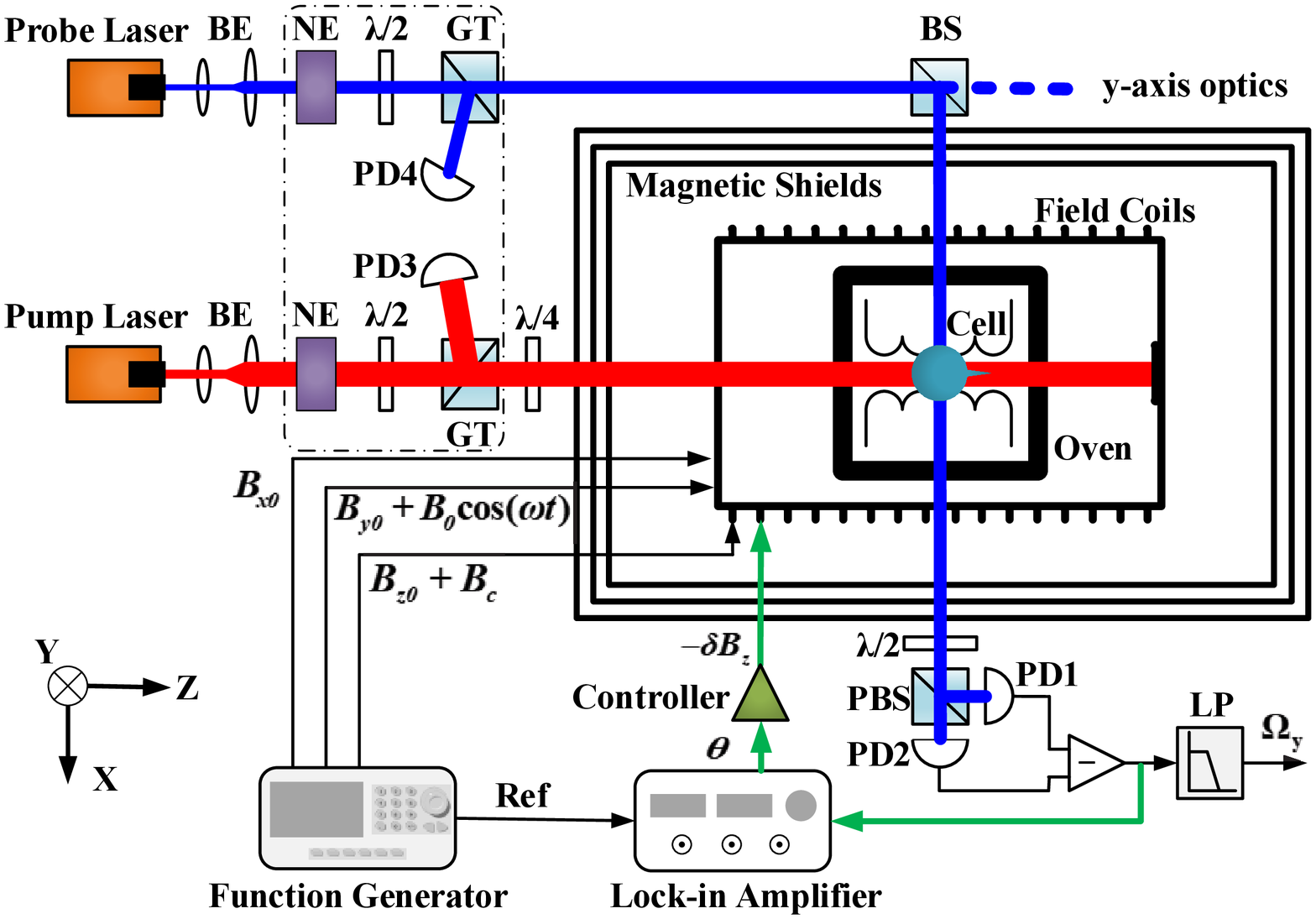}
\caption{\label{fig1} Schematic of the experimental apparatus. A circularly polarized pump laser propagating along $z$ axis is used to polarize atoms. A linearly polarized probe laser is split by a beam splitter (BS) to measure the transverse polarization component in the $x$ axis and $y$ axis respectively, facilitating dual-axis output. The $y$-axis optics are similar to that along $x$ axis. A homemade laser intensity electrocircuit sampling the signal of photodiodes (PD3 and PD4) stabilizes the intensities of the pump laser and probe laser using the noise eaters (NE) as actuators. The output is analyzed by a lock-in amplifier to extract the transverse oscillating field response and then feed back to control the compensation point. The low pass filter (LP) removes the high-frequency response to extract the rotation sensing signal. Beam expander (BE); Glan-Taylor polarizer (GT); Polarization beam splitter (PBS).}
\end{center}
\end{figure}

The experiment is performed in a dual-axis K-Rb-$^{21}$Ne comagnetometer which is used for rotation sensing and depicted in Fig.~\ref{fig1}. A 10-mm-diameter spherical cell made from GE180 aluminosilicate glass, containing a droplet of natural abundance Rb with a small admixture of K, 3 atm $^{21}$Ne (70\% isotope enriched), and 60 Torr N$_{2}$ for quenching, is used. The cell is placed in a boron nitride ceramic oven and heated to 190$^{\circ}$C by a homemade 129-kHz ac electrical heater. At the operating temperature, the Rb vapor density of about 6\texttimes10$^{14}$\ cm$^{-3}$ is obtained and the density ratio of K to Rb is approximately 1:80. The cell and oven are surrounded by three-layer $\mu$-metal cylindrical magnetic shields and a set of three-axis magnetic coil system. The magnetic coil system consists of two pairs of saddle coils along $x$ and $y$ axis respectively and two pairs of Lee-Whiting coils with different constants along $z$ axis. The transverse coils and the larger constant longitudinal coil (Z1 coil) driven by a function generator are used to compensate the residual magnetic fields within the innermost shield, provide transverse field modulation and set the compensation point. The smaller constant longitudinal coil (Z2 coil) driven by the controller is used to finely control the compensation point. K atoms are optically pumped along $z$ axis by a 38-mW pump laser, centered on K D1 resonance line. Rb atoms are polarized by the K atoms through spin exchange interaction, then they hyperpolarize the $^{21}$Ne atoms \cite{happer2001,Hyper}. The transverse polarization of Rb atoms is measured by optical rotation of a linearly polarized probe laser using about 1 mW and tuned by 0.3 nm to the blue side of Rb D1 resonance line \cite{Faraday}.

\begin{figure}[b]
\begin{center}
\includegraphics[width=7.5cm]{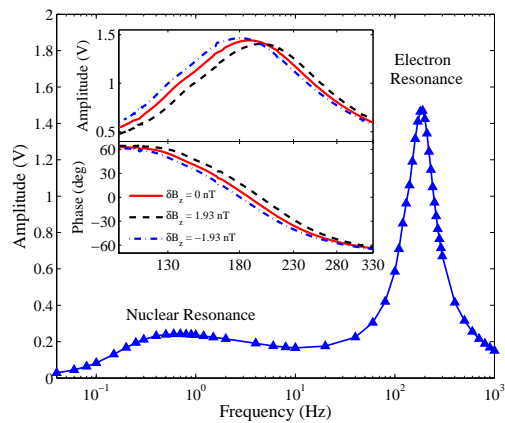}
\caption{\label{fig2} The frequency response of the K-Rb-$^{21}$Ne comagnetometer at the compensation point to oscillating fields along the $y$ axis. The electron resonance is clearly separated from the nuclear resonance. The inset shows the amplitude-frequency and phase-frequency response around the electron resonance. The resonance frequency and phase of electron scale with the shift of the compensation point. }
\end{center}
\end{figure}

Here we firstly investigate the frequency response of K-Rb-$^{21}$Ne comagnetometer to transverse oscillating magnetic field. A series of magnetic fields with different frequencies and the same peek-to-peek amplitude of 0.15 nT are produced along $y$ axis. The frequency response measured by $x$-axis optics is shown in Fig.~\ref{fig2}. The electron resonance separates far away from the nuclear resonance, which is different from that of K-$^{3}$He comagnetometer \cite{kornack2002}. The discrepancy is owing to the large field from electron magnetization in the Rb-$^{21}$Ne comagnetometer. Under the normal comagnetometer operation, a bias magnetic field ${B_c} =  - {B_n} - {B_e}$ called the compensation point is applied parallel to the direction of pump beam to cancel the field from nuclear magnetization ${B_n}=\frac{8}{3}\pi {\kappa _0}{M_n}P_z ^n$ and the field from electron magnetization ${B_e}=\frac{8}{3}\pi {\kappa _0}{M_e}P_z ^e$. Therein the atomic spins experience an effective field equal to their own magnetization. $\kappa _0$ is the enhancement factor arising from the overlap of the alkali metal electron wavefunction and the noble gas nucleus \cite{K}. $M_n$ and $M_e$ are the magnetization of nuclear spins and electron spins corresponding to full spin polarization, which are proportional to the atom number density. $P_z ^n$ and $P_z ^e$ are the $z$-axis components of nuclear spin polarization ${\bf{P}}^{\bf{n}}$ and electron spin polarization ${\bf{P}}^{\bf{e}}$, respectively. The enhancement factor $\kappa _0$ for Rb-$^{21}$Ne pair is about 5 times larger than that for K-$^{3}$He pair \cite{RbNe,KHe}. Meanwhile, the density of Rb atom is about one order higher than that of K atom in the typical K-$^{3}$He comagnetometer. Therefore, the electron spins experience a much larger magnetic field in the K-Rb-$^{21}$Ne comagnetometer, which shifts the electron resonance frequency to high frequency and keeps the spin exchange relaxation to be not completely eliminated. An analytic explanation for this phenomenon is presented hereafter.

The behavior of the comagnetometer can be described by a set of coupled Bloch equations for ${\bf{P}}^{\bf{n}}$ and ${\bf{P}}^{\bf{e}}$.  For small transverse excitations of the spins, the angles of polarization vectors ${\bf{P}}^{\bf{n}}$ and ${\bf{P}}^{\bf{e}}$ with respect to the $z$ axis are small enough, so that we approximately assume the longitudinal polarization components $P_z^n$ and $P_z^e$ as constants \cite{kornack2005,li2016}. We focus on the electron resonance, so an oscillating field $B_0$cos$({\omega}t)$ is applied along $y$ axis, whose frequency is much higher than nuclear resonance frequency. Ignoring the minor impact of nuclear spins on high-frequency response of electron spins, the oscillating electron spin polarization measured by $x$-axis optics can be approximated to the following illuminating form \cite{SM},

\begin{flalign}\label{eq:pexw}\begin{split}
P_x^e\left( t \right) =& \frac{{{\gamma _e}{B_0}P_z^e}}{2}\bigg[ {\frac{{R_{tot}^e\cos \left( {\omega t} \right) + \left( {Q\omega  - {\gamma _e}{B_z^a}} \right)\sin \left( {\omega t} \right)}}{{{{\left( {Q\omega  - {\gamma _e}{B_z^a}} \right)}^2} + R{{_{tot}^e}^2}}}} \\
& { + \frac{{R_{tot}^e\cos \left( { - \omega t} \right) - \left( {Q\omega  + {\gamma _e}{B_z^a}} \right)\sin \left( { - \omega t} \right)}}{{{{\left( {Q\omega  + {\gamma _e}{B_z^a}} \right)}^2} + R{{_{tot}^e}^2}}}} \bigg],
\end{split}&
\end{flalign}
where $\gamma _e$ is the gyromagnetic ratio of electron spins and $R_{tot}^e$ is the total relaxation rate for electron. $Q$ is the nuclear slowing-down factor and is a function of the electron polarization \cite{Q}. $B_z^a$ is the total effective magnetic field along $z$ axis experienced by the electron.

The dynamics response takes the form of two overlapping Lorentzian curves centered at $ \pm {\gamma _e}{B_z^a}/Q$ and both of the curves have a component with an absorptive lineshape that is in-phase with the oscillating field and a component with a dispersive lineshape that is 90$^\circ$ out-of-phase with the oscillating field. As the electron resonance frequency is much larger than the linewidth which is shown in Fig.~\ref{fig2}, the counter-rotating response centered at $ -{\gamma _e}{B_z^a}/Q$ can be ignored \cite{lu}. With the compensation point enforced, the amplitude-frequency response $\left| {P_x^e}{\left( \omega  \right)} \right|$ and phase-frequency response $\theta{\left( \omega  \right)}$ can be simplified to the intuitive form $\left| {P_x^e}{\left( \omega  \right)} \right|{\rm{ = }}\frac{{{\gamma _e}{B_0}P_z^e}}{{2\sqrt {{{\left( {Q\omega {\rm{ + }}{\gamma _e}{B_e} - {\gamma _e}\delta {B_z}} \right)}^2} + R{{_{tot}^e}^2}} }}$ and $\theta{\left( \omega  \right)} {\rm{ = }} - {\rm{arctan}}\big( {\frac{{Q\omega  + {\gamma _e}{B_e} - {\gamma _e}\delta {B_z}}}{{R_{tot}^e}}} \big)$, where $\delta {B_z}$ is the magnetic field with respect to the compensation point along $z$ axis. When the residual magnetic field inside the shields and light shift field compensated by Z1 coil, $\delta {B_z}$ approaches 0 at the compensation point.

We fit the electron resonance at the compensation point in Fig.~\ref{fig2} by amplitude-frequency response $\left| {P_x^e}{\left( \omega  \right)} \right|$. The resonance frequency ${\omega _0} =  - {\gamma _e}{B_e}/Q$ is about $188 \pm 0.8$ Hz and the resonance linewidth $\Delta \omega {\rm{ = }}R_{tot}^e/Q$ is about $50\pm 1.8$ Hz. For typical electron polarization $P_z^e \approx $ 60\% in our experiment, we find $Q \approx $ 7.5. Then the electron magnetization $B_e$ is about $-$50 nT, which is approximately an order of magnitude larger than that in K-$^{3}$He comagnetometer. The total electron relaxation rate $R_{tot}^e$ is about 2356 s$^{-1}$, which is highly suppressed from the spin-exchange rate of 5.5\texttimes10$^{5}$\ s$^{-1}$. Although experienced a large magnetic field in the K-Rb-$^{21}$Ne comagnetometer, the alkali metal electron spins are still operated in the near SERF regime \cite{Happer1977,Romalis2002}.

\begin{figure}[b]
\begin{center}
\includegraphics[width=7.5cm]{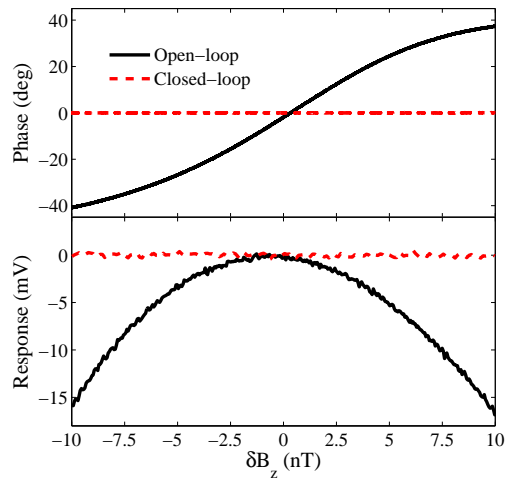}
\caption{\label{fig3} Demodulated phase at driving frequency $\omega _0$ and signal of the comagnetometer response to the shift of compensation point scanning from $-10$ nT to $10$ nT in the open-loop scheme and closed-loop scheme. The output offset at the compensation point is set to zero. By stabilizing the demodulated phase at electron resonance, the comagnetometer is free from the drift of longitudinal magnetic field in the closed-loop scheme.  }
\end{center}
\end{figure}

From amplitude-frequency response $\left| {P_x^e}{\left( \omega  \right)} \right|$ and phase-frequency response $\theta{\left( \omega  \right)}$, we can see that the electron resonance frequency and phase scale with the shift of compensation point $\delta {B_z}$ and the result measured by $x$-axis optics is shown in the inset of Fig.~\ref{fig2}. This phenomenon inspires the closed-loop control of the compensation point by locking the electron resonance. To accomplish this, a feedback control system has been incorporated in our apparatus which is depicted in Fig.~\ref{fig1}. A field modulated at the electron resonance frequency $\omega _0$ is applied along $y$ axis, the signal is read out by the $x$-axis optics and demodulated by the lock-in amplifier with $cos({\omega _0}t)$. The demodulated phase $\theta {\rm{ = arctan}}\left( {{\gamma _e}\delta {B_z}/R_{tot}^e} \right)$ is fed through the controller. The controller compares the demodulated phase with the initial electron resonance phase, then powers the Z2 coil to keep the electron resonance constant.

The performance of the closed-loop control system is evaluated by slowly scanning $\delta {B_z}$ field using Z1 coil from $-10$ nT to $10$ nT around the compensation point \cite{SM}. The demodulated phase at driving frequency $\omega _0$ and signal of the comagnetometer is recorded by a National Instrument 24-bit data acquisition system and summarized in Fig.~\ref{fig3}. The residual magnetic fields $B_x$ and $B_y$, rotations $\Omega _x$ and $\Omega _y$, light shift fields $L_x$ and $L_z$ can not be zeroed completely, which will introduce a $\delta {B_z}$ dependence to the signal \cite{jiang2017},
\begin{flalign}\label{eq:pexBz}\begin{split}
P_x^e\left( {\delta {B_z}} \right) \approx &\frac{{ - \gamma _e^2P_z^e\delta {B_z}}}{{R{{_{tot}^e}^2}{B_n}}}\Big[ {\big( {{B_x} + \frac{{Q{\Omega _x}}}{{{\gamma _e}}} + {L_x}} \big)} \delta {B_z}{\rm{ + }} {{L_z}{B_x}} \\
&  { + \frac{{R_{tot}^e{B_y}}}{{{\gamma _e}}} - \frac{{{B_c}{\Omega _x}}}{{{\gamma _n}}} + \frac{{QR_{tot}^e{\Omega _y}}}{{\gamma _e^2}} + {B_c}{L_x}}  \Big],
\end{split}&
\end{flalign}
where $\gamma _n$ is the gyromagnetic ratio of nuclear spins.

\begin{figure}[t]
\begin{center}
\includegraphics[width=7.5cm]{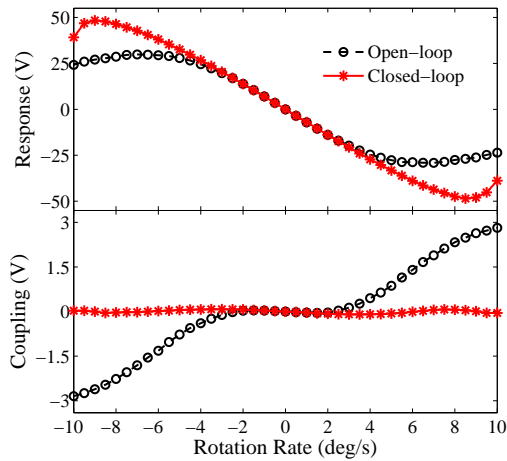}
\caption{\label{fig4} The dual-axis comagnetometer response to rotation rate inputting along $y$ axis operated in the open-loop scheme and closed-loop scheme. Top: The sensitive response $P_x^e$ measured by $x$-axis optics. Bottom: The coupling response $P_y^e$ measured by $y$-axis optics. The closed-loop control of the compensation point could extend the linear measurement range and suppress the cross-talk error of comagnetometer. }
\end{center}
\end{figure}

 When operated in the open-loop scheme, the demodulated phase at driving frequency $\omega _0$ response to $\delta {B_z}$ scanning is consistent with the theoretical expression $\theta {\rm{ = arctan}}\left( {{\gamma _e}\delta {B_z}/R_{tot}^e} \right)$, while the signal of the comagnetometer drifts with $\delta {B_z}$ scanning and the profile is a quadratic function corresponding to Eq.~(\ref{eq:pexBz}). When operated in the closed-loop scheme, $\delta {B_z}$ is real-time compensated by feedback electronics, thus the demodulated phase and the signal of the comagnetometer remain constant. With the closed-loop control of the compensation point, the comagnetometer is unaffected by the fluctuation of the longitudinal residual magnetic field.

Next we consider the effect of fluctuation of the nuclear magnetization. $\delta {B_z}$ is constantly drifting on account of drifting nuclear spin polarization. When the transverse excitations are large, ${\bf{P}}^{\bf{n}}$ will precess a large angle away from $z$ axis and the constant assumption of $P_z^n$ is not positive. We use a rotating platform with an accuracy of 0.001 deg/s to provide the transverse excitations. The $y$ axis of the comagnetometer is mounted vertically and aligned with the rotating axis of the platform. When inputting $\Omega _y$, the signal measured by $x$-axis optics is defined as the sensitive response and shown in Eq.~(\ref{eq:spx}), meanwhile, the signal measured by $y$-axis optics arising from the detune of compensation point $\delta {B_z}$ is defined as the coupling response and shown in Eq.~(\ref{eq:spy}).
\begin{equation}\label{eq:spx}
P_x^e(\Omega _y) = \frac{{ - {\gamma _e}P_z^eR_{tot}^e{\Omega _y}/{\gamma _n}}}{{R{{_{tot}^e}^2}{\rm{ + }}\gamma _e^2\big[ {\delta {B_z}^2{\rm{ + }}{{\left( {{\Omega _y}/{\gamma _n}} \right)}^2}} \big]}},
\end{equation}
\begin{equation}\label{eq:spy}
P_y^e(\Omega _y) = \frac{{ - \gamma _e^2P_z^e\delta {B_z}{\Omega _y}/{\gamma _n}}}{{R{{_{tot}^e}^2}{\rm{ + }}\gamma _e^2\big[ {\delta {B_z}^2{\rm{ + }}{{\left( {{\Omega _y}/{\gamma _n}} \right)}^2}} \big]}}.
\end{equation}

 The experimental result is shown in Fig.~\ref{fig4}. When the comagnetometer operated in the open-loop scheme, with the inputting rotation rate increasing, the equilibrium angle between ${\bf{P}}^{\bf{n}}$ and $z$ axis increases, making $\delta {B_z}$ gradually deviate from the compensation point. The nonzero $\delta {B_z}$ decreases the response of the sensitive axis and arises a severe cross-talk response in the coupling axis. When the the comagnetometer operated in the closed-loop scheme, nonzero $\delta {B_z}$ is real-time canceled by the feedback electronics, leaving the compensation point tuned all the time, which can improve the scale factor linearity, enlarge the measuring range and suppress the cross-talk error of the comagnetometer. The minor fluctuation of the coupling response around zero in the closed-loop scheme can be further removed by optimizing the control algorithm and parameters.

The limited measuring range in the closed-loop scheme is restricted by the characteristics of the atom spins, which is given by ${\Omega _{\max }} =  \pm {\gamma _n}R_{tot}^e/{\gamma _e}$. While the rotation uncertainty per unit bandwidth is given by $\delta \Omega  = {\gamma _n}R_{tot}^eQ/\left( {{\gamma _e}nV} \right)$, where $n$ is the density of alkali metal atoms and $V$ is the measurement volume\cite{kornack2005}. In the future, we should balance the discrepancy between high sensitivity and large measuring range of the SERF comagnetometer, furthermore, a closed-loop detection method is still needed to investigate to extent the measuring range.

In conclusion, we have demonstrated a real-time closed-loop control method to stabilize the compensation point by locking the electron resonance in the K-Rb-$^{21}$Ne comagnetometer, which has not been previously investigated, either theoretically or experimentally. The technique presented here could improve the systematic stability, enlarge the linear measuring range, and suppress cross-talk error of the comagnetometer, which will be important for precision metrology applications using the SERF comagnetometer, particularly for rotation sensing as gyroscopes and fundamental physics tests of spin interactions beyond the standard model \cite{SM, tests2017}.

We would like to thank Wenfeng Wu and Yan Yin for useful discussions. This work was supported by the National Key R\&D Program of China under grants Nos. 2016YFB0501600, 2016YFA0301500, NSFC under grants Nos. 61773043, 61473268, 61503353, 11434015, 61835013, SPRPCAS under grants Nos. XDB01020300, XDB21030300.

\bibliography{reference}

\end{document}